# SU-8 clamped CVD graphene drum resonators

Sunwoo Lee[1, a)], Changyao Chen[2, a)], Vikram V. Deshpande[3], Gwan-Hyoung Lee[2,4], Ilkyu Lee[3], Michael Lekas[1], Alexander Gondarenko[2], Young-Jun Yu[5], Kenneth Shepard[1], Philip Kim[3], James Hone[2, b)]

[1]*Department of Electrical Engineering, Columbia University, New York, New York 10027, USA.*
[2]*Department of Mechanical Engineering, Columbia University, New York, New York 10027, USA.*
[3]*Department of Physics, Columbia University, New York, New York 10027, USA.*
[4]*Samsung-SKKU Graphene Center (SSGC), Suwon, Gyeonggi 440-746, Korea.*
[5]*Creative Research Center for Graphene Electronics, Electronics and Telecomumnications Research Institute (ETRI) Daejeon, 305-700, Korea.*

Graphene mechanical resonators are the ultimate two-dimensional nanoelectromechanical systems (NEMS) with applications in sensing and signal processing [1]. While initial devices have shown promising results [2-4], an ideal graphene NEMS resonator should be strain engineered, clamped at the edge without trapping gas underneath [5], and electrically integratable. In this letter, we demonstrate fabrication and direct electrical measurement of circular SU-8 polymer-clamped chemical vapor deposition (CVD) graphene drum resonators. The clamping increases device yield and responsivity, while providing a cleaner resonance spectrum from eliminated edge modes. Furthermore, this resonator is highly strained, indicating its potential in strain engineering for performance enhancement.

Graphene, a two-dimensional material consisting of a single layer of carbon atoms, has enormous mechanical strength and exceptional electrical properties [6,7]. Both of these superior qualities can be utilized in the field of nanoelectromechanical systems (NEMS) [2,3]. The high stiffness and low mass density of graphene provide high resonant frequencies, its ultrahigh strength allows large strain tuning, and its large carrier mobility enables self-signal amplification [8,9] for electrical transduction [10,11,12]. Owning to their small mass and large mechanical compliance, graphene NEMS resonators have great potential for sensitive mass, force, and charge detection [10].

Most studies of graphene NEMS resonators, and all those employing electrical transduction, utilize a doubly clamped geometry, whereby a suspended rectangular graphene strip (typical length scale of 1-5μm) is supported at the ends by metal electrodes. These devices show resonant frequencies of 10's to 100's of MHz and room-temperature quality factors of ~100 [3,10]. Recent studies have shown that, when the suspended graphene is fully clamped on all sides, the room-T quality factor can exceed 1000 in devices with diameters above 13 μm [4]. Furthermore, these drum resonators show a highly regular series

---

a) Sunwoo Lee and Changyao Chen contributed equally to this work
b) Author to whom correspondence should be addressed. Electronic mail: jh2228@columbia.edu

of resonant frequencies, without spurious edge modes. However, these devices were read out optically, and integrating fully clamped devices into an electrical transduction scheme is challenging. Previously demosntrated methods for fabricating fully clamped graphene NEMS have deposited graphene onto pre-patterned holes. Performing additional processing steps such as lithography for shaping the graphene and applying electrodes is highly problematic on already-suspended devices. Moreover, the graphene membranes trap gas in the holes beneath [5]; it can require days for this trapped gas to escape when the samples are placed in vacuum for testing. Attempts have been made to resolve this issue, including making a hole at the center of the resonator [13]. However, such methods risk the mechanical integrity of the resonator.

In this work, we demonstrate a novel technique to fabricate fully clamped graphene resonators that avoids both of these problems. SU-8 polymer is used to fully clamp the graphene membrane from the top, and the SU-8/graphene heterostructure is released together in the final processing step, eliminating any enclosed space and eliminating any processing on suspended graphene. We have found that the addition of SU-8 polymer in graphene resonator fabrication does not degrade the electrical quality of graphene. The clamping allows for greater fabrication yield and contributes to achieving better spectral purity. Furthermore, increased resonant frequency due to SU-8 induced strain is observed.

Fig. 1 shows a schematic of the fabrication process. We use insulating substrates such as highly resistive silicon (>10 kΩ cm) or fused silica in order to minimize radio frequency (RF) crosstalk during electrical measurements. On such substrates, local gates and alignment marks (1/30 nm of Ti/Au) are patterned at the wafer scale using deep-UV lithography (ASML 300C DUV stepper). A layer of silicon oxide is then grown by plasma-enhanced chemical vapor deposition (PECVD) on top of the electrodes to bury the local gate. To improve adhesion of graphene to the substrate during the subsequent transfer step, chemical mechanical polishing (CMP) is used to smooth the capping PECVD oxide surface down to about 0.4nm in roughness. The structure of the substrate is shown in Fig. 1.(a).

To further demonstrate the applicability of this technique for wafer scale integration, we use large-grain CVD graphene in this study, which has electrical and mechanical properties close to those of mechanically exfoliated graphene [14,15]. Graphene is grown on copper foil[16] at 1070 °C with low (1sccm) methane flow to obtain large grain size (>100 μm) growth. After the graphene is grown, we spin-coat a layer of poly(methyl methacrylate) (PMMA) and then press a thick layer of polydimethyl siloxane (PDMS) elastomer on top in order to increase mechanical stability during transfer [17]. After wet-etch of the copper (Transene, APS100), the graphene /PMMA/PDMS stack is then pressed against the aforementioned substrate and heated to 170 °C to allow slow peeling of the PDMS. Finally, the PMMA is stripped in acetone, leaving with only graphene on the wafer, as shown in Fig. 1.(b). The graphene sheet is then patterned into a series of strips above the buried metal gates with electron beam lithography (EBL, Nano Beam nB4) and oxygen plasma etch (Plasma Etch PE-50). Source and drain electrodes (1/15/50 nm of Ti/Pd/Au) are then patterned again using EBL, and lift-off (Fig. 1.(c)). After each step, the PMMA is stripped in acetone, followed by immersion in chloroform at 75 °C for 30 minutes to clean the remaining PMMA residue [18].

Once the source and drain electrodes are defined, a SU-8 layer of 2 μm thickness is spin-coated and patterned by EBL to be the same dimension as the graphene strips, except for the presence of hole in the center. After the SU-8 patterning, the sample is hard-baked at 170 °C for 30 minutes to induce cross-linking in the SU-8. This step also induces strain on the graphene membrane as the SU-8 clamp shrinks during the hard-bake. Lastly, the sample is immersed in buffered oxide etchant and dried in critical point dryer, in order to suspend the SU-8 clamped graphene heterostructure (Fig. 1.(e)). We used scanning electron microscopy (SEM) to examine the suspended structure, as shown in Fig. 1.(f).

The SU-8 clamping improves the mechanical rigidity of the suspended structure, which is critical for bringing NEMS resonators to practice. In particular, clamping the graphene edges allows us to decrease gate-to-graphene distance and to apply larger DC gate voltage, both of which directly increase the electro-mechanical coupling and therefore the readout efficiency: the signal current is proportional to $1/z^3$, where z is the distance between the suspended graphene and the gate electrode, owing to enhancement of both the electrostatic force and the gate coupling [9]. We have fabricated SU-8 clamped resonators with graphene-gate gap as small as 50 nm, and diameter-to-gap size ratio of 40. We find that the yield of SU-8 clamped drum resonators with 50 nm gap size is greater than 70%, while the yield for standard doubly clamped resonators with 50 nm gap size is less than 10%. With the SU-8 support, we can apply large $V_g^{DC}$ up to 10V for 1.5μm diameter drums with 50nm gap while 3V is enough to collapse doubly-clamped resonators with the same dimension, demonstrating how this performance enhancement is only feasible through the SU-8 clamp. In fact, because of this improved transduction and structural rigidity for high biasing, we have been able to apply the direct detection scheme to relatively low quality graphene resonators, even the ones with the mobilities only ~500 $cm^2$/V-s.

The samples are measured in a vacuum ( < $3\times10^{-5}$ Torr) probe station at room temperature. We leave the devices in vacuum for several hours before measurement in order to remove moisture and some chemical residues. To examine the effect of SU-8 on electrical performance, we measured the gate voltage-dependent resistance of two types of suspended graphene samples on the same chip: ones with SU-8 clamps and the others without. The result is shown in Fig. 2. We observe that both the graphene with and without SU-8 are slightly p-doped by few volts, probably due to the chemical process associated with the transfer. The field effect mobilities of typical SU-8 clamped resonators are about 3000 $cm^2$/V-s, which we have routinely observed on non-SU-8 clamped resonators at room temperature, indicating that SU-8 clamping does not greatly degrade the electrical properties of graphene. The highest mobilities we observed for both resonators with SU-8 and without SU-8 are about 6000 $cm^2$/V-s.

A previously described resonant channel transistor scheme [9] was used to transduce mechanical resonances. The circuit diagram is shown in Fig. 3.(a): with the source grounded, DC bias $V_d^{DC}$ is applied to the drain, and a combined DC bias $V_g^{DC}$ and RF signal from a vector network analyzer (VNA, Agilent E5072C) are applied to the gate through a bias tee (Mini-circuits ZFBT-4GW+). A second bias tee is used to separate the DC and RF drain currents, and the RF output signal is passed through an amplifier (Miteq AU•1447•BNC) before being fed back into the VNA. Fig. 3(b) shows the measured transmission coefficient $S_{21}$ as a function of drive frequency for a 4μm diameter graphene drum resonator at $V_g^{DC}$ = 8.6V, $V_d^{DC}$ = -500mV, and drive power of -40dBm. The device shows a clear resonance peak near 48 MHz, with a quality factor of ~60, which is about factor of four lower than previously reported for circular drum resonators of the same diameter [4]. This decrease in Q maybe be due to SU-8 residue, displacement current induced dissipation [12], or wrinkling caused by non-uniform strain.

Fig. 4.(a) shows the measured $S_{21}$ as function of both frequency and $V_g^{DC}$. With increasing $V_g^{DC}$, the resoant frequency first descrases (spring constant softening), then increases (spring constant hardening). The spring constant softening is due to nonlinear electrostatic interaction and the hardening arises from elastic strentching of the graphene membrane [19]. This non-monotonic behavior, indicative of high strain, was observed in many other samples we tested. Applying a contiuum mechanical model to fit the experimental result, with only built-in strain and total mass as fitting parameters, we find a built-in strain of 0.0023. This is more than one order of magnitude larger than the room-temperature values for typical graphene resonators previously reported [2,4], indicating that the SU-8 processing imparts tension to the membrane. The effective mass density is 6.3 times of the pristine graphene, mainly due to the resist residues culmulated over processing steps. Comparable amounts of resist residue have been previously observed in graphene exposed only to PMMA.

A primary benefit of the clamped geometry is the purity of the mechanical resonance. Doubly-clamped graphene resonators can exhibit complex patterns of resonant modes due to edge effects [20] and to the vibration of the metal contacts [10], which are also suspended due to the fast diffusion of BOE at the graphene/oxide interface [21]. As an example, Fig. 4.(b) shows the measured resonance of regular doubly clamped graphene resonator fabricated with same condition except for the SU-8 clamps. It is clearly seen that the strip resonator possesses multiple $V_g^{DC}$ - independent metal contact resonances contrary to the clean resonance spectrum for SU-8 clamped drum resoantors shown in Fig. 4.(a). We observe similar patterns of multiple resonances in ~20% of doubly-clamped devices, but none in the SU-8 clamped devices. We note that the SU-8 block is also suspended, but its mechanical resonance can be pushed far away from the graphene resonancce by simply changing the dimensions and geometry of the SU-8 clamp.

The enhanced strain increases the resonant frequency. The highest fundamental resonant mode we measure is 260 MHz for 1.5μm diameter drum, with gate bias of 6V, drain bias of 1V, and drive power of -55dBm at room temperature, as shown in Fig. 5.(a). This value is almost one order of mangitude larger than that of devices with the same size made by deposition across pre-patterned holes [4], indicating that the SU-8 processing induces subtantial strain to the membrane. In this device, we measure the strain directly using atomic force microscope (AFM) nanoindentation [7]. The measured force-displacement curve is shown in Fig. 5.(b). The slope of the force-displacement curve at low deflection gives a strain of 0.004, which is even higher than that of the 4 μm diameter drum shown above. Similar strain engineering methods have been previously reported using SU-8 in a doubly-clamped geometry [22], using thermal processing to tension circular graphene oxide membranes [13], and using chemically modified graphene film [23]. However, our work demonstrates for the first time a strain-engineered graphene drum resonator with full electircal integration. Further improvement for larger and more uniform strain using different SU-8 hard-bake temperature or the clamping geometry, coupled with cleaner processing steps should be able to deiliver electrically integrated GHz graphene resonator with F·Q product exceeding $10^{12}$.

In summary, we demonstrated a novel fabrication process for fully clamped graphene drum resonators using SU-8 polymer. The additional SU-8 does not degrade the electrical quality of CVD graphene, and increases the mechanical rigidity of the suspended structure to enhance the electro-mechanical responsivity of the resonator. The high yield fabrication result indicates the possibility of reliable suspended graphene resonators at wafer scale, while the frequency enhancement paves path toward strain engineering on graphene resonator using SU-8.

The authors like to thank Victor Abramsky and Jerry D. Smith for critical discussions. Fabrication was performed at the Cornell Nano-Scale Facility, a member of the National Nanotechnology Infrastructure Network, which is supported by the National Science Foundation (Grant ECS-0335765), and Center for Engineering and Physical Science Research (CEPSR) Clean Room at Columba University. The authors acknowledge the support by Qualcomm Innovation Fellowship (QInF) 2012 and AFOSR MURI FA9550-09-1-0705.

FIG. 1. (a-e) Schematic of fabrication processes of SU-8 clamped graphene drum resonator. (f) false-color SEM image of graphene drum resonator with 30 degrees tilt angle. Scale bar: 4 μm. (Inset: same device with 60 degrees tilt angle showing the gap between the SU-8 clamped graphene and local gate).

FIG. 2. Resistance ($R$) as function of applied DC gate voltage ($V_g$) for SU-8 clamped drum resonator (Black) and non-clamped graphene strip resonators (Red).

FIG. 3. Measurement setup and typical resonance plot for a drum resonator with 4 μm diameter. (a) Circuit-level schematic of direct detection measurement setup. (b) Typical resonance data taken at room temperature (black dots) and Lorentzian fit (red), both in logarithmic scale.

FIG. 4. Color plot of $S_{21}$ as a function applied gate voltage ($V_g$) and frequency for (a) SU-8 clamped graphene drum resonator and (b) graphene strip resonator. The yellow curve in (a) is a continuum mechanical model fit as described in the main text.

FIG. 5. Resonance and strain measurements of highly strain graphene resonator. (a) $S_{21}$ measurement showing the resonance frequency of 260MHz (b) Strain measurement with AFM nanoindentation showing load-displacement characteristic curve for SU-8 clamped graphene. The linear regime modeling indicates that the pre-strain of 0.4% due to SU-8.

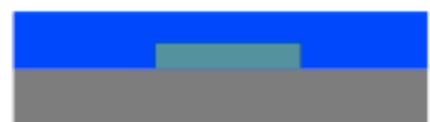
(a) Local Gate with SiO₂ Cladding & CMP

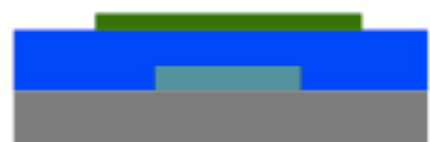
(b) Graphene Transfer & Patterning

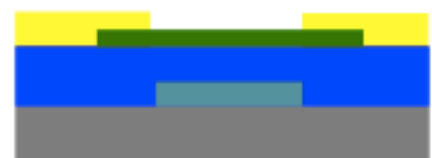
(c) Source & Drain Electrodes

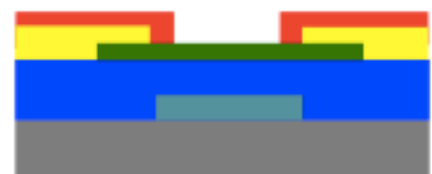
(d) Circular Clamping Using SU-8

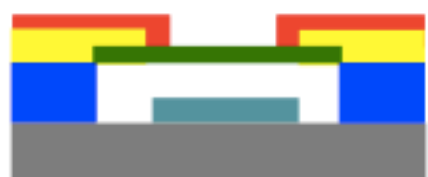
(e.1) BOE Release

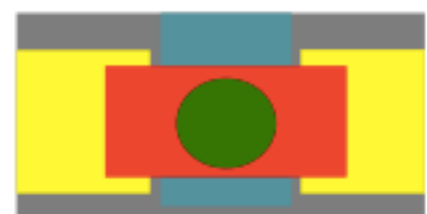
(e.2) BOE Release *top-view*

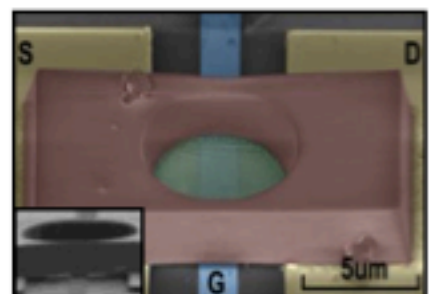
(f) False Color SEM

| | | | |
|---|---|---|---|
| 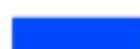 | PECVD SiO₂ | 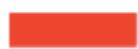 | SU-8 |
| 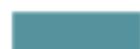 | Local Gate | 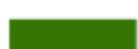 | Graphene |
| 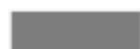 | Silicon | 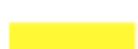 | Source/Drain |

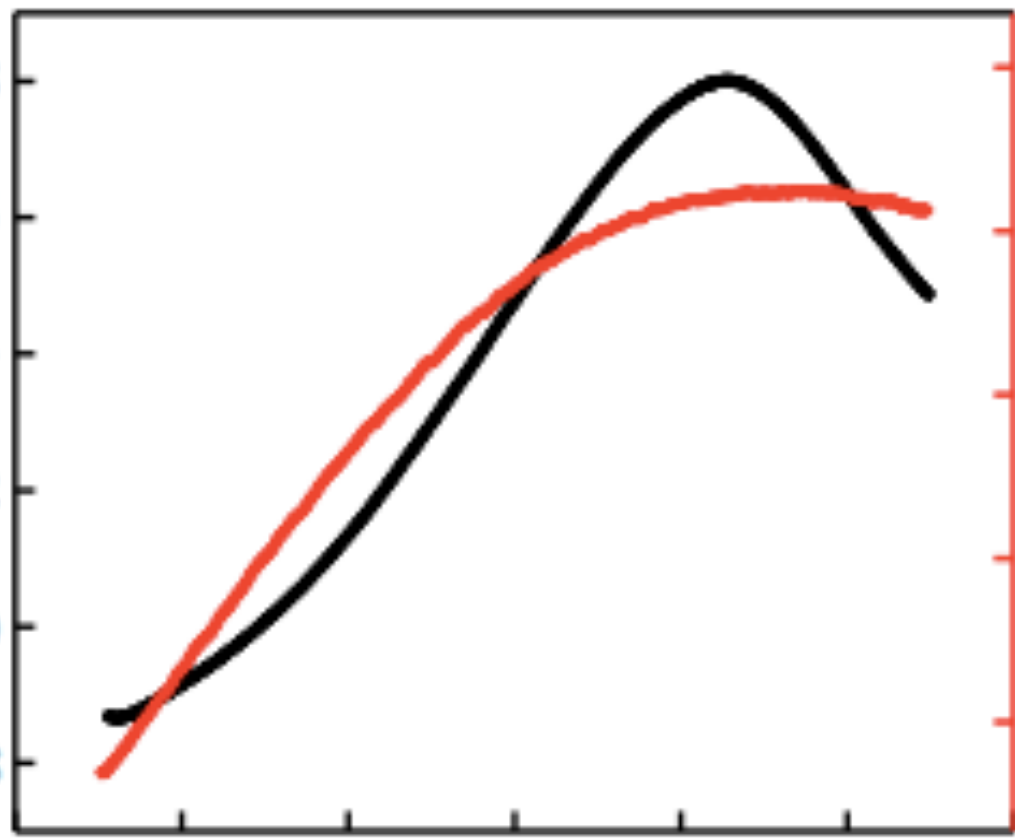

(a) 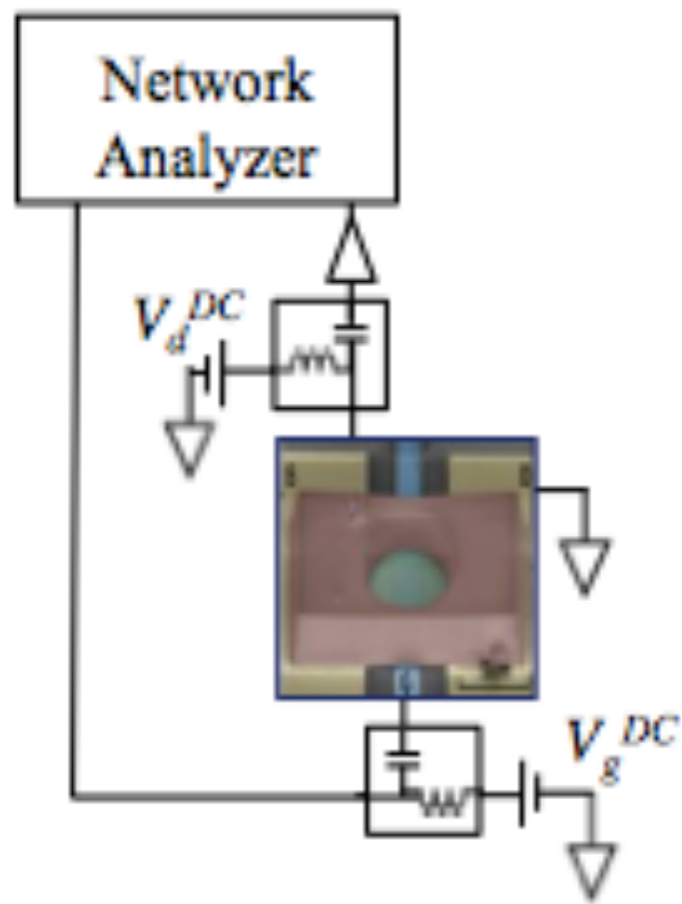
(b) 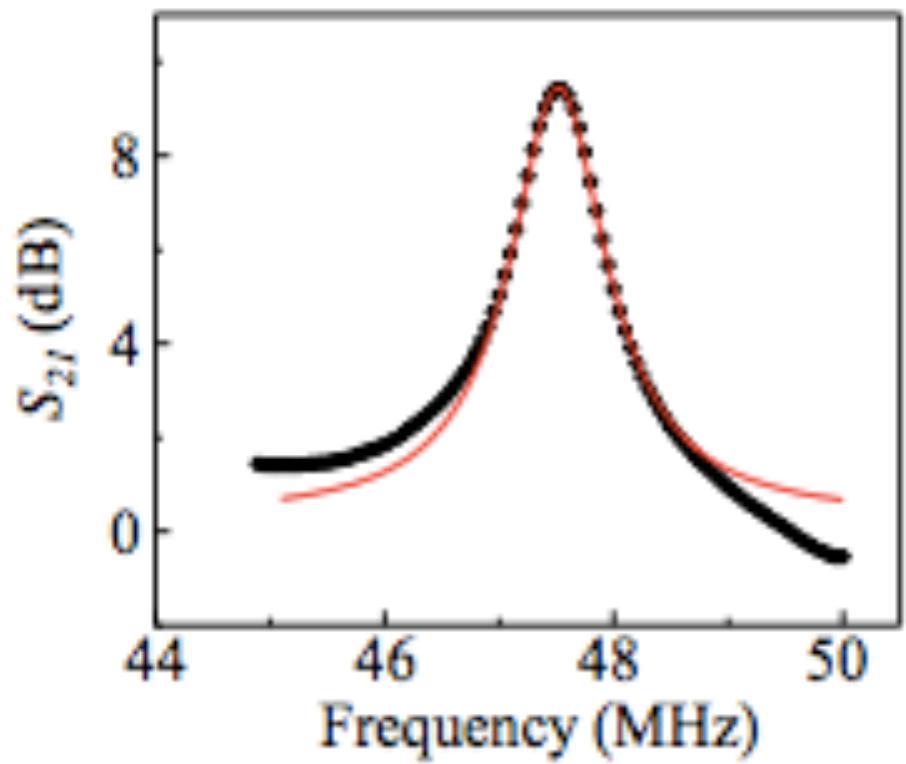

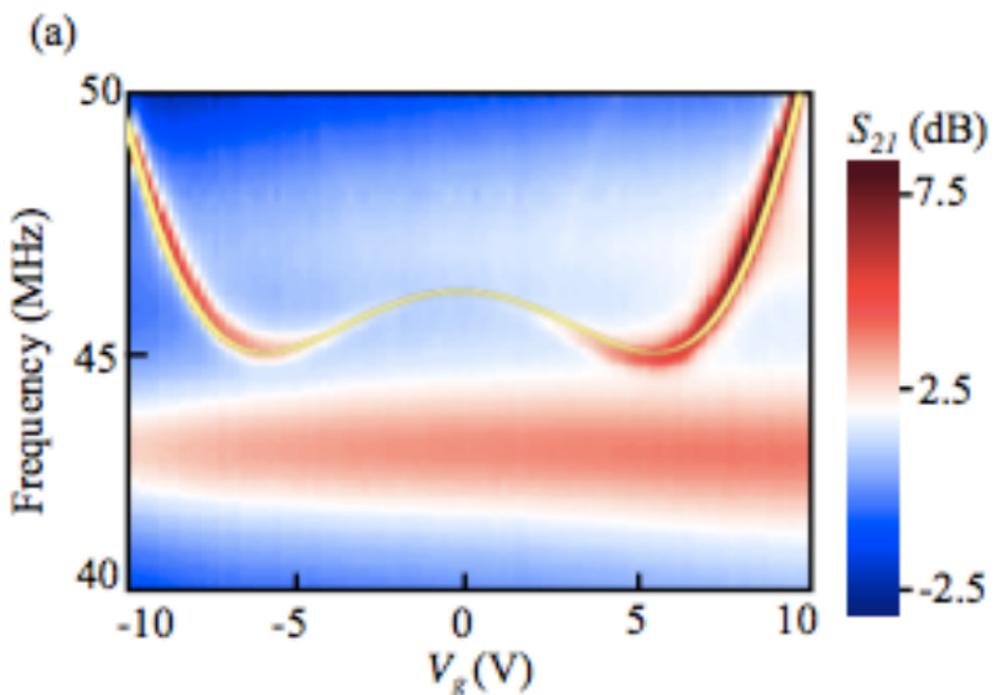

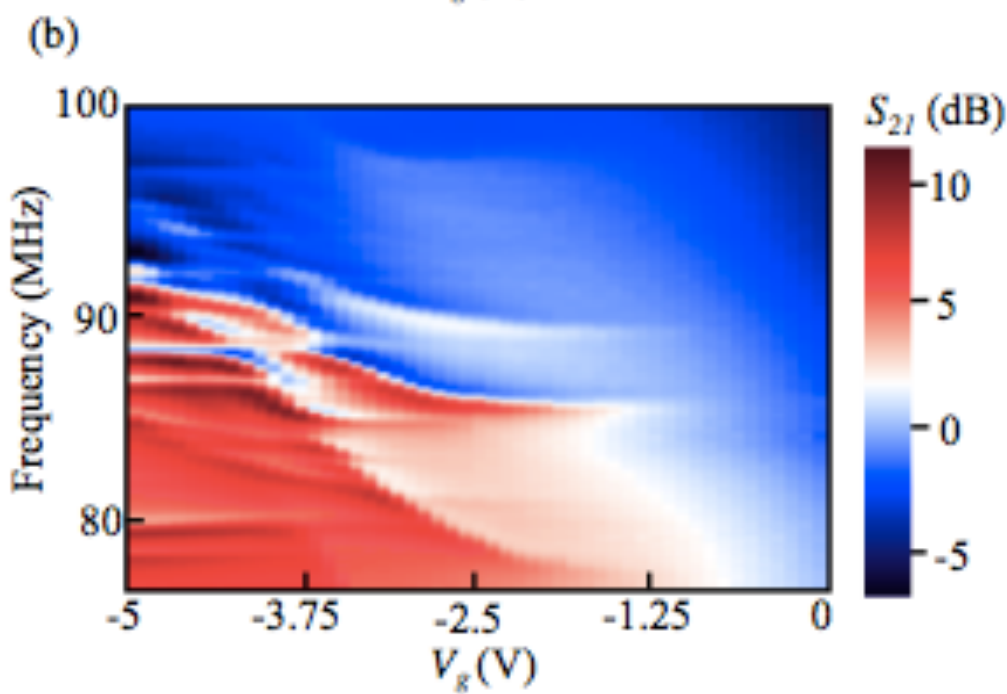

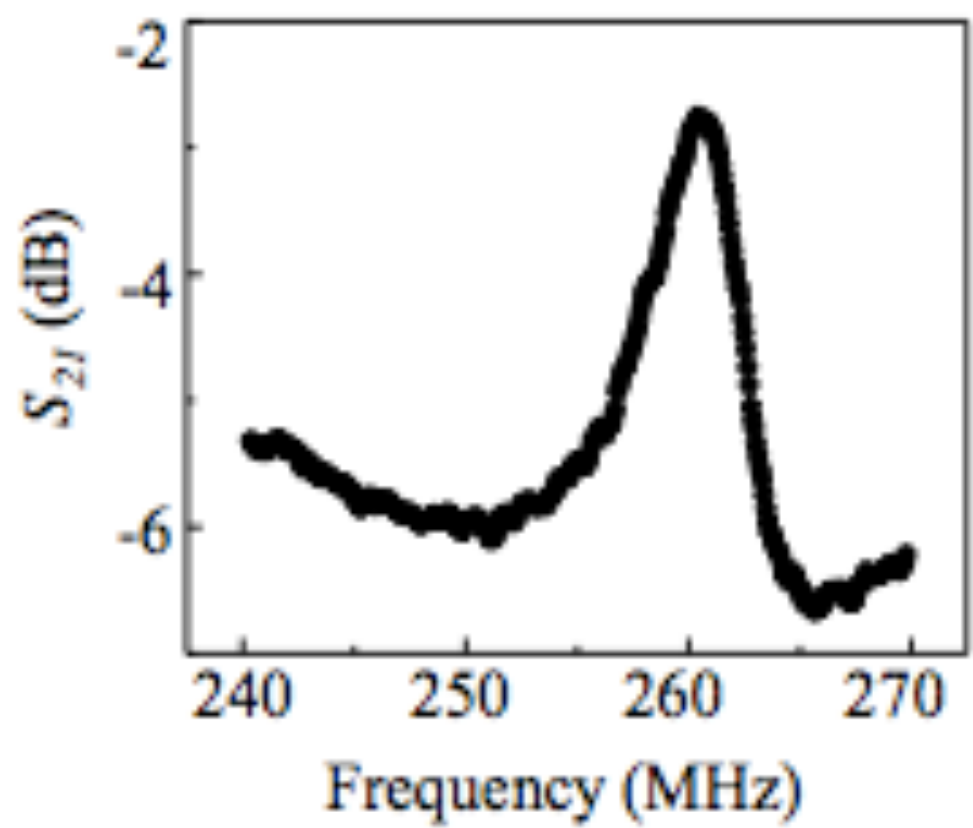 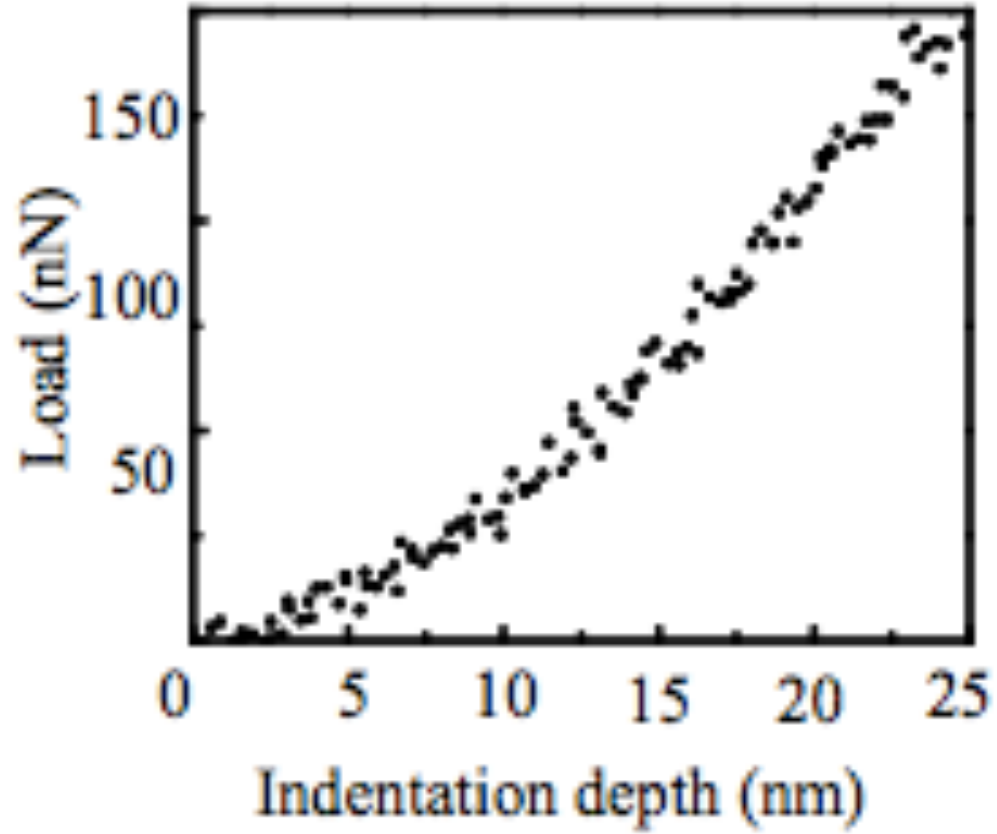